\documentclass[aps,prl,reprint,groupedaddress]{revtex4-1}

\usepackage{pbox}
\usepackage{bm}
\usepackage{amsmath}
\usepackage{graphicx}
\usepackage{dcolumn}
\usepackage[version=4]{mhchem}
\usepackage{url}
\begin{document}

\title{Finding reaction pathways with optimal atomic index mappings}


% \CJKfamily{gbsn}

%\thanks{}
\author{Deb Sankar De}
\affiliation{Department of Physics, Universit\"at Basel, Klingelbergstr. 82, 4056 Basel, Switzerland}
\author{Marco Krummenacher}
\affiliation{Department of Physics, Universit\"at Basel, Klingelbergstr. 82, 4056 Basel, Switzerland}
\author{Bastian Schaefer}
\affiliation{Department of Physics, Universit\"at Basel, Klingelbergstr. 82, 4056 Basel, Switzerland}
\author{Stefan Goedecker}
\email{stefan.goedecker@unibas.ch}
\affiliation{Department of Physics, Universit\"at Basel, Klingelbergstr. 82, 4056 Basel, Switzerland}

\begin{abstract}
Finding complex reaction and transformation pathways, involving many intermediate states, is in general not possible on the DFT level
with existing simulation methods due to the very large number of required energy and force evaluations. This is due to a large extent to 
the fact that for complex 
reactions, it is  not possible to determine which atom in the educt is mapped onto which atom 
in the product. Trying out all possible atomic index mappings is not feasible because of the factorial increase in the number of possible 
mappings. By using a penalty function that is invariant under index permutations, we can bias the 
potential energy surface in such a way that it obtains the characteristics of a structure seeker whose global minimum is the product. By 
performing a Minima Hopping based 
global optimization on this biased potential energy surface we can rapidly find 
intermediate states that lead into the global minimum. Based on this information 
we can then extract the full reaction pathway. We first demonstrate for a benchmark system, namely $LJ_{38}$ that our method allows to 
preferentially find 
intermediate states that are relevant for the lowest energy reaction pathway and that we therefore need a much smaller number of 
intermediate states than previous methods to find the lowest energy reaction pathway.
Finally we apply the method to two real systems, \ce{C60} and \ce{C20H20}  and show that the found reaction pathway 
contains valuable information on how the system can be synthesized.
\end{abstract}

%\keywords{}

\maketitle

The indistinguishability of atoms manifests itself in various ways.
In a chemical reaction the indistinguishability means that in principle any atom of a certain type in the educt can 
be mapped onto any atom of the same type in  the product. If the system undergoing the chemical reaction consist of 
$M_1$ atoms of type 1, $M_2$ atoms of type 2, etc, then there are consequently $M_1! M_2! .... $ different possible 
mappings. For simple reactions it is often obvious which atom in the educt is mapped onto which atom of the product.
In more complicated chemical reactions or similar processes such as phase or shape transformations in nano-particles 
the best mapping can however not be foreseen. Actually in such a context many atomic index mappings may exist which  
lead to different low barrier reaction pathways.

Determining the transition states of a 
reaction/transformation  
is of great importance in chemistry, physics and materials sciences.
A large number of methods have therefore been proposed to solve this 
problem as reviewed for instance by Reiher~\cite{simm2018exploration}. 
In the so-called two-sided 
methods~\cite{NEB,baer,ghasemi} one has not only 
to provide the educt and the product but also the correct atomic index mapping, 
i.e. it has to be known beforehand which atom in the educt is 
mapped on which atom in the product. In principle 
one could of course perform such a transition state search for all 
possible atomic index mappings. In practice this is of course 
prohibitive for large systems because the factorial 
increase in the number of possibilities.

In the so-called one sided methods~\cite{dimer,abashkin1994transition} one finds a transition state that is in a certain sense closest 
to the initial guess. Once arrived at the transition state one can find the 
adjacent minima by performing two simple local geometry optimizations starting from a point to the left and a point to 
the right of the saddle point 
along the line of negative curvature one can thus find the two minima that are connected by this saddle point. 
In this way one obtains automatically the correct index mapping, but it is not guaranteed that the two local minima found in this way are 
the desired educt and product.

There is  a method that would allow  in principle to study any chemical reaction or transformation,
namely Molecular Dynamics (MD). In this approach there is no index mapping problem since any atom of the educt is mapped 
onto its corresponding atom in the product via its trajectory. However, since most chemical reactions are rare events, the number of MD time steps is prohibitive.  
Meta dynamics~\cite{Laio}  methods speed up processes that are slow on the time 
scale of a typical MD time step. They are driven by collective variables that ``push" the system in a desired direction.
The standard collective variables can target certain phases or general structural motifs, but they can not target a single 
configuration. A typical collective variable that is used to drive chemical reactions in a desired  direction is for instance 
a certain bond length, whose value in the product is known and different from the value in the educt. But it is very likely that there are other stable 
configurations that have the same bond length and so the collective variable can drive the system into a configuration that is different from the product.
In the field of protein folding, certain averages of the dihedral angles that give rise to Ramachandran plots are frequently 
used as collective 
variable. Again, it is obvious that many configurations of the protein may give rise to the same average angles. 
Also very recently proposed  collective variables, such as approximate 
entropies~\cite{parrinello} do not allow to target a single configuration.

Temperature accelerated dynamics~\cite{voter} is another way to speed up rare events that would be impossible to observe with standard 
MD. This class of methods has the advantage that it does not need collective variables, but on the other hand 
it is again not possible to steer the reaction pathways in a certain desired direction.
High temperature MD trajectories are used in a similar spirit in the 
TSSCDS method~\cite{Vazquez2018} to detect saddle points.

Minima hopping guided path search~\cite{mhgps} (MHGPS) is another unbiased method to explore the entire Potential Energy Surface (PES) and to find 
a large number of local minima and transition states. 
Essentially the same information can be obtained by combining basin hopping~\cite{BH}  with 
an eigenvector following approach~\cite{Doye1997} to find saddle points.
So all these  methods provide in principle the information required to construct any reaction pathway with a correct index mapping, but do not allow for steering. 
 For small systems exploring the entire PES is feasible and of interest to find a more or less complete 
 set of minima together with the saddle points connecting them. For large systems such an approach is however numerically too expensive 
 and it is advantageous to concentrate 
 on a subset of minima and saddle points that are of interest in a certain context.
Transition path sampling~\cite{bolhuis2002transition} is yet another method that allows for the calculation of multiple reaction pathways 
between given 
initial and final configurations. However again the initial atomic index matching has to be known beforehand. 

In this paper we will use configurational distances derived from fingerprints that are invariant under atomic index permutations~\cite{sadeghi}  
as a driving force towards the desired final configuration. In contrast to the standard collective variables, this fingerprint 
can uniquely identify a single configuration. The fingerprint distance is zero if and only if the configuration is 
exactly identical to the desired final configuration up to translations, rotations and 
index permutations. Along the reaction pathway the distance varies continuously until it 
vanishes at the final configuration. Adding the distance as a penalty $P({\bf R}_1,...,{\bf R}_M)$ to the true 
PES $E({\bf R}_1,...,{\bf R}_M)$ gives a `biased'
PES $\tilde{E}({\bf R}_1,...,{\bf R}_N) = E({\bf R}_1,...,{\bf R}_N)$ + w $P({\bf R}_1,...,{\bf R}_N)$. 
The parameter $w$ is in this context a suitable chosen weight and ${\bf R}_1,...,{\bf R}_N$ are the cartesian coordinates of the $N$ atoms 
in the system. The penalty is based on a fingerprint obtained from the eigenvalues of an overlap matrix $S$ describing the
configuration~\cite{sadeghi}. Each pair of atoms (i , j) in the configuration gives rise to a block of 
matrix elements given by 
$ S_{i,j} = \int G_{\alpha}({ \bf r} - {\bf R}_i)  G_{\beta}({ \bf r} - {\bf R}_j) d{\bf r} $.
The indices $\alpha$ and $\beta$ specify the type of the orbitals. We used Gaussian 
type orbital of $s$, $p_x$, $p_y$ and $p_z$ character. 
Hence  $S_{i,j}$ is a 4 by 4 matrix block. 
Denoting by ${\bf V}^p$ the vector containing the sorted eigenvalues of the overlap matrix 
characterizing  configuration p and by ${\bf V}^q$ the analogous vector of configuration q, 
the distance $D_{p,q}$ between the two configurations was defined as  
$$ D_{p,q} = \sqrt{ \sum_j (V_j^p-V_j^q)^2} $$ 
in the original publication~\cite{sadeghi}.
This definition of the distance can lead to discontinuties in the first derivative with 
respect to the atomic positions when eigenvalues cross. Such discontinuities make 
a numerical optimization of the penalty term in the biased potential energy more difficult.
We therefore removed these discontinuities by introducing smeared out 
eigenvalues  ${\bf \tilde{V}}_p$ given by
$$ \tilde{V}_i^p = \frac{  \sum_l V_l^p \exp(-\frac{1}{2} \left( \frac{(V_l^p-V_i^p)}{a} \right)^2   )}
				   {\sum_l \exp(-\frac{1}{2} \left( \frac{(V_l^p-V_i^p)}{a} \right)^2   )} $$
The smeared eigenvalues do also cross, but at the crossing they have identical derivatives.  
Our regularized distance serves then as the penalty function:
$$ P = \sqrt{ \sum_j (\tilde{V}_j^p-\tilde{V}_j^q)^2} $$ 
All derivatives of this function are continuous. 
Adding the gradient with respect to the atomic positions of this penalty term to the physical 
forces, gives biased 
forces that drive the system from the present configuration $p$ towards the desired final configuration $q$. 
These forces are invariant with respect 
to index permutations.

\begin{figure*}[!htbp]
\includegraphics[width=0.9\textwidth,angle=0]{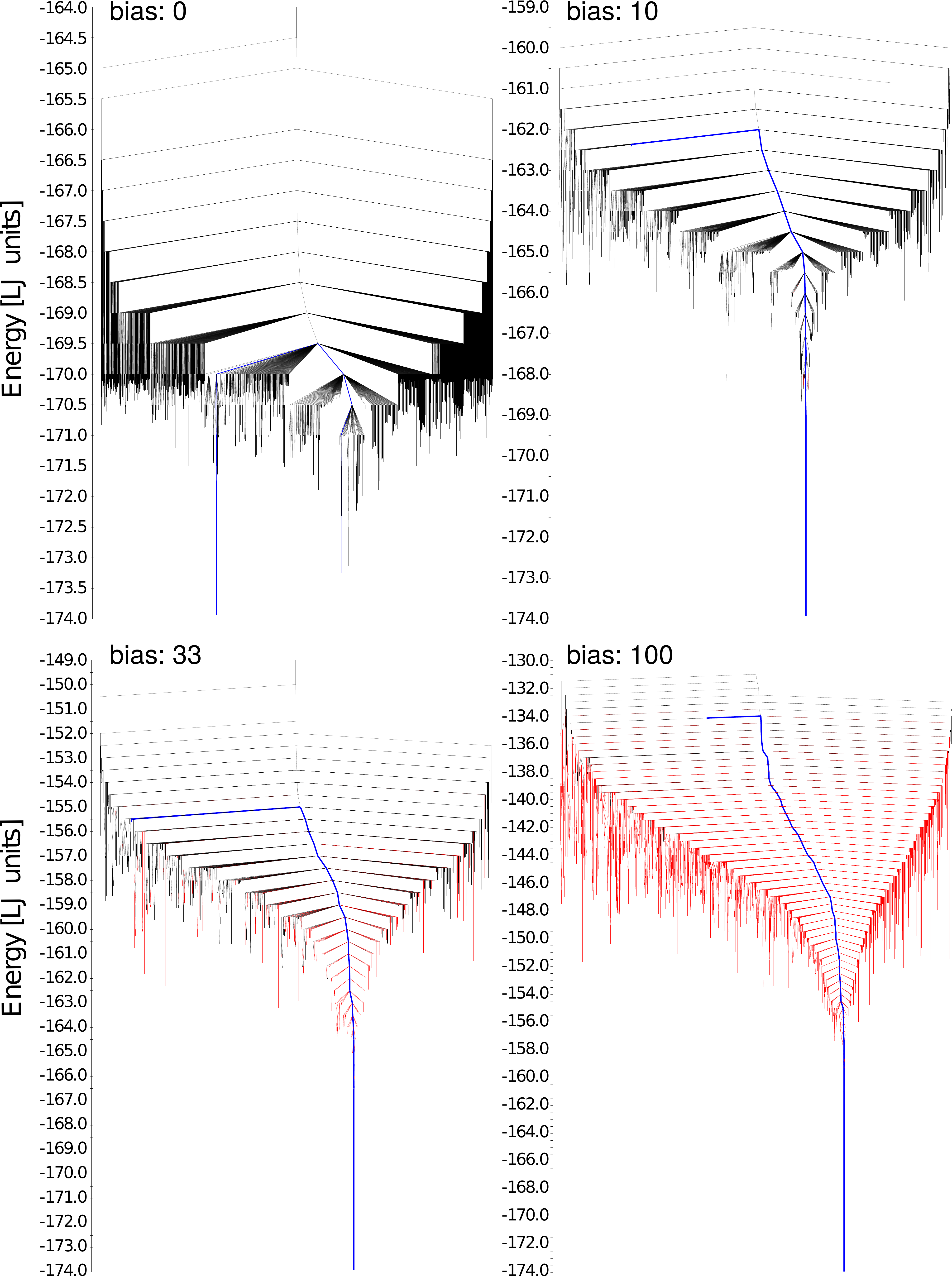}
\caption{ Disconnectivity graphs of the $LJ_{38}$ cluster with different biased PESs. The
red minima indicate fcc like structures and the blue path shows the transition from the 
second lowest icosahedral to the global fcc minimum. The graph was constructed with the 
disconnectionDPS software~\cite{dpsSoft}. } 
\label{fig:LJ38trees}
\end{figure*}

In this work we concentrate on complex reaction pathways, where the system has to cross a substantial number of 
saddle points.  By adding the penalty we transform the original PES $E$ into a biased one, denoted by $\tilde{E}$.
With the right choice of the parameter $w$, this biased PES has the appearance of 
a PES of a structure seeker whose global minimum is the desired final 
state  as can be seen from the disconnectivy graphs in Fig.~\ref{fig:LJ38trees})~\cite{becker1997topology,dpsSoft}. 
This means that the downhill barriers (the barriers that one has to overcome when one 
hops from one minimum into another one with lower energy) are reduced but have  
not disappeared. Therefore a 
local optimization of the biased PES is not sufficient to find the final configuration. 
A global optimization is required. 
However, we exploit the fact that for a structure seeker it is easy to find the global minimum with the minima hopping method. We start from 
an initial configuration (educt) and then visit with the minima hopping methods consecutive intermediate states until we 
find the ground state. Because of the stochastic nature of minima hopping different intermediate states can be visited in different runs.
With a reasonable choice for the weight of the penalty all these intermediate configurations will have low energies and are therefore 
physically possible intermediate states that are accepted during a Minima Hopping run. This means that one can find with this approach not only one reaction pathway but all 
physically relevant low energy reaction pathways.

Since the hops from one local minimum to the next in Minima Hopping are 
based on MD~\cite{MDBEP,softening} followed by local 
geometry optimizations~\cite{Schaefer2015},
the identity of individual atoms of the same type can be traced back for any hop and one obtains in this way the correct mapping of 
the atomic indices for the entire complex reaction pathway. The MD trajectories cross barriers and so one 
could take some configurations along the MD trajectory as a starting point for a one sided saddle point search. The MD trajectory can however cross over several barriers 
within one hop of a MH run and in such a case it is not clear which point along the MD trajectory should be chosen as the starting point for the saddle point search. 
We therefore used a recursive variant of the freezing string method~\cite{freezestring} as described in  
reference~\cite{mhgps} to connect the sequence of accepted local minima by saddle points. The method described so far will be referred to as Biased Minima Hopping 
Guided Pathway Search (BMHGPS) in the following.

We will first apply our new method to a benchmark system, the  
Lennard Jones cluster with 38 atoms ($LJ_{38}$), 
for which the lowest energy reaction pathway from the lowest energy icosahedral structure to the fcc ground state is most likely known, since it was studied  previously~\cite{doye1999,neirotti2000phase,mandelshtam2006multiple,sehgal2014phase}.
Finding this transformation pathway is quite difficult since it requires  a complete overall rearrangement of all the atoms in the system~\cite{doye1999}. 

To find the biased PES that has the strongest structure seeker 
character, we used four different weights $w$ of the penalty function ($w=0, 10, 33, 100$). Choosing $w=0$ gives back the unbiased MHGPS method, which was included for comparison. For each weight 100 different MH runs were performed. As reference structure the fcc  global minimum was taken in order to bias the minimization towards it.

\begin{figure}[!htbp]
\includegraphics[width=1.0\columnwidth,angle=0]{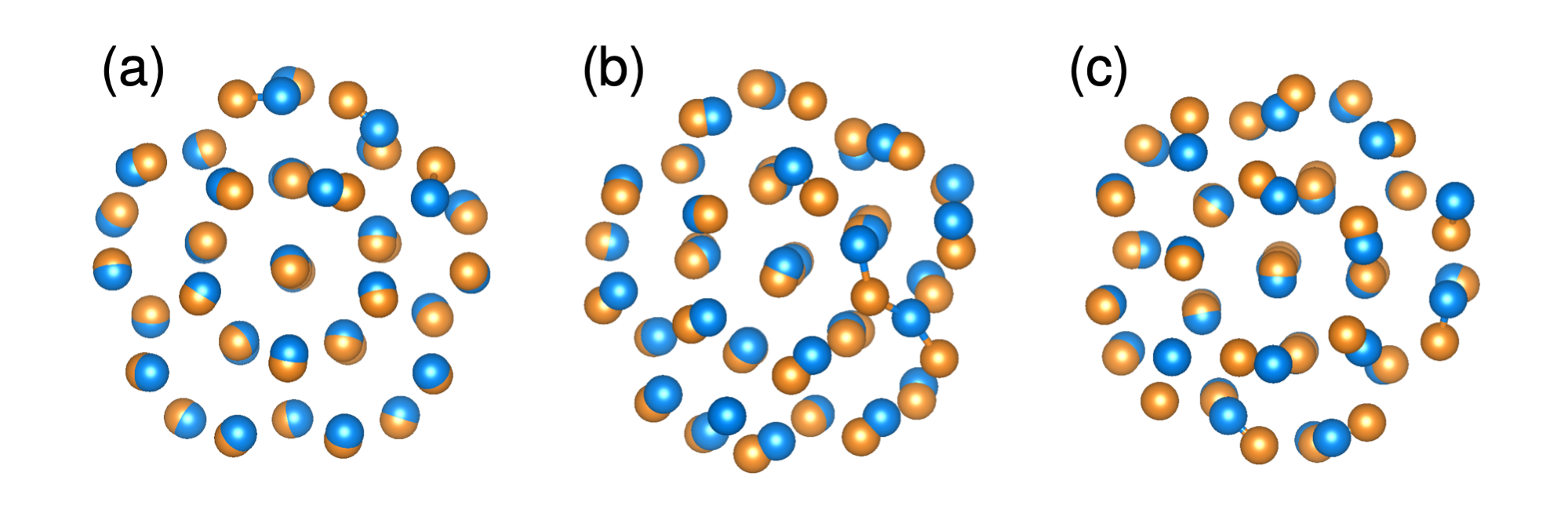}
\caption{The icosahedral equilibrium configuration on the physical PES (orange) compared to a biased PES (blue) for a bias of (a) w=10, (b) w=33 and (c) w=100. 
} 
\label{fig:posall}
\end{figure}

As can be observed in the disconnectivity graphs in Fig.~\ref{fig:LJ38trees} the typical double funnel structure of the PES disappears by adding the penalty function to the energy. For all non-zero weights, the disconnectivity graph has the shape of 
a structure seeker for which it is quite easy to find the global minimum. The more the PES is biased the more fcc like minima can be observed. On the non-biased PES, no 
fcc structure except the global minimum can be seen 
(shown in this case in blue and not red since it represents the transformation pathway).
For large weights fcc like structures start to dominate. 
Choosing even larger weights would result in a glassy landscape of fcc like structures where it would again be more difficult to find a pathway into the global minimum. In addition too large weights will destroy a faithful mapping between 
the local minima on the biased PES and the true physical PES. 
Fig.~\ref{fig:posall} illustrates this mapping for the icosahedron starting structure.  It can be seen that the displacement of the surface atoms due to the bias is 
larger than for the core atoms. Following the entire transformation obtained by our method indeed shows that it 
starts by some kind of surface premelting which then propagates towards the centre.

\begin{figure}[!htbp]
\includegraphics[width=1.0\columnwidth,angle=0]{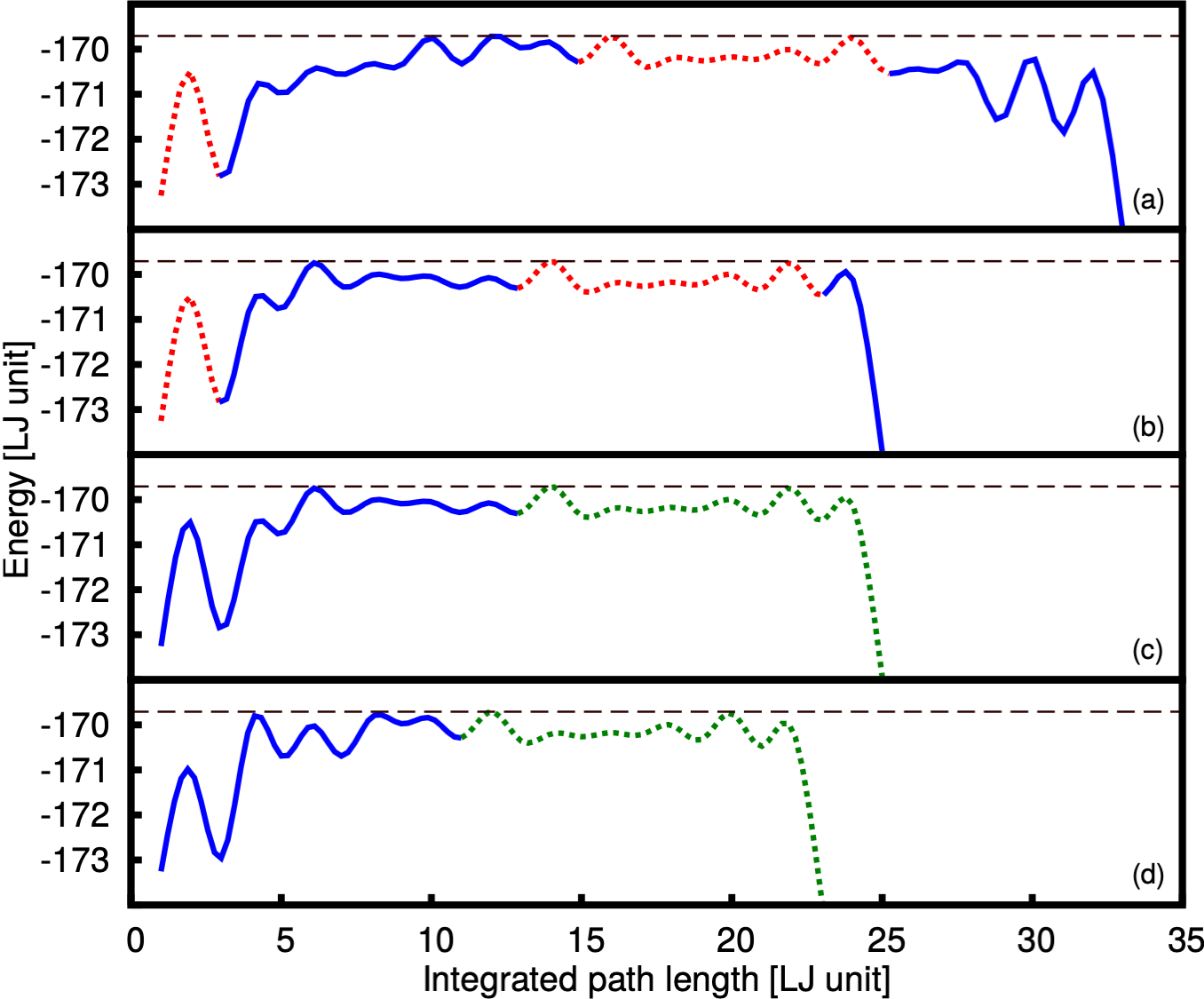}
\caption{ BMHGPS pathways found for the $LJ_{38}$ cluster with different bias strengths (a) $w=100$, (b) $w=33$, (c) $w=10$ and (d) $w=0$ connecting the second lowest icosahedral minimum with the global fcc minimum. The dashed horizontal lines indicates the highest energy along the lowest known pathway~\cite{doye1999,doye1999_2}.
The sections indicated by dashed lines are shared by pathways obtained with different weights. The pathways found with a weight of $w=10$ and $33$ are identical.
}
\label{fig:LJ38path}
\end{figure}

All the MH runs were stopped once the global minimum was found.
The number of distinct minima that were visited before the global minimum was found was greatly reduced by the bias.  Averaging over 100 runs we visited on average 
363 minima without any bias, 19 for $w=10$, 10 for $w=33$ and 16 for $w=100$.
In order to obtain reaction pathways the accepted local minima from the MH run on the biased PES are transformed back to the physical PES by a local geometry optimization. 
Reaction pathways were then generated as described in reference~\cite{mhgps}. All our pathways shown in  Fig.~\ref{fig:LJ38path} have a highest energy barrier of -169.709 in LJ energy units which is identical to the lowest highest barrier height found in previous studies~\cite{doye1999,doye1999_2}. The reaction pathway 
found in these two studies was however based on a data set of 140 000 saddle 
points whereas we could obtain this result thanks to the bias with a much 
smaller data set. We used all the saddle points obtained in the 100 runs 
and collect in this way 3238 saddle points for $w=10$, 2068 for $w=33$ and 
4183 for $w=100$. Without a bias we used a data set of similar size, as the previous studies, namely 83 000 saddle points.
This reduction in the number of required saddle points is due to the fact that our biasing allows us to preferentially find 
the intermediate states and saddle points in the region of configurational space that is in between 
the educt and the product. Those states are the relevant ones for our wanted reaction pathway from the educt to the product.

Because of the significant efficiency gains obtained by BMHGPS, it can also be applied on the Density Functions Theory (DFT) level and we 
will use it to find highly complex reaction pathways for \ce{C60} and \ce{C20H20}. The DFT calculations were performed with the BigDFT 
code~\cite{bigdft} 
using dual space Gaussian pseudopotentials~\cite{Willand2013}. 
An initial pathway  was obtained  
with the  self-consistent
charge density functional tight-binding method (SCC-DFTB) 
as implemented in DFTB+ ~\cite{aradi2007dftb+} with s and p atomic orbitals for carbon. These pathways were then further refined with BigDFT calculations. 

Several  MD  studies of \ce{C60} 
formation failed to find a route from high-energy
structures to the Buckyball ground state, presumably because the necessary time scales
are not accessible to MD simulations~\cite{ballone1990simulated,chelikowsky1991nucleation,chelikowsky1992jr,jing1992nucleation,yi1993reactivity,xu1994tight,marcos1997thermal}. 
Only a semiclosed \ce{C60}  pseudocage was found
after a simulation time of 250 ns  using a classical Brenner potential and a adpative temperature MD method~\cite{maruyama1998molecular}.

\begin{figure}[!htbp]
\includegraphics[width=1.0\columnwidth,angle=0]{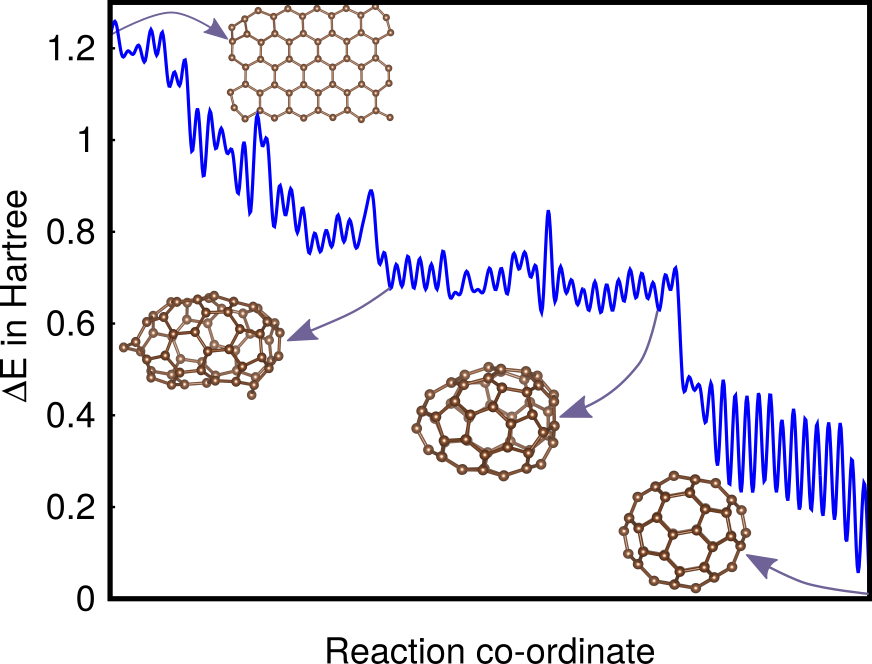}
\caption{Typical low-barrier pathway of \ce{C20H20} from the graphene flake to the fullerene. } 
\label{fig:C60graph}
\end{figure}

Using our new method, we were able to find a reaction pathway from an intial structure that was a planar graphene flake to the Buckyball ground state.
Good weights $w$ are in the range of 0.2 to 0.3 which means that the energy 
difference between the ground state and the first Stone-Wales defect is increased  
0.06 Ha to 0.08 Ha and the energy difference between the initial flake and 
the ground state is increased from 1.21 Ha to 1.97 Ha.
The low energy reaction pathway shown in Fig. ~\ref{fig:C60graph} is based on a data set 
of 9000 saddle points that were obtained from 4 MH runs that were stopped once 
the global minimum was found. Even though the barriers along the pathway are quite 
high, there is a clear tendency to lower the energy as one approaches the ground state.
This is also true for other representative  pathways that we have found. 
This explains why \ce{C60} can be synthesized in 
a one-step procedure by graphite vaporization at high temperatures~\cite{C60synthesis}. 
  
\begin{figure}[!htbp]
\includegraphics[width=1.0\columnwidth,angle=0]{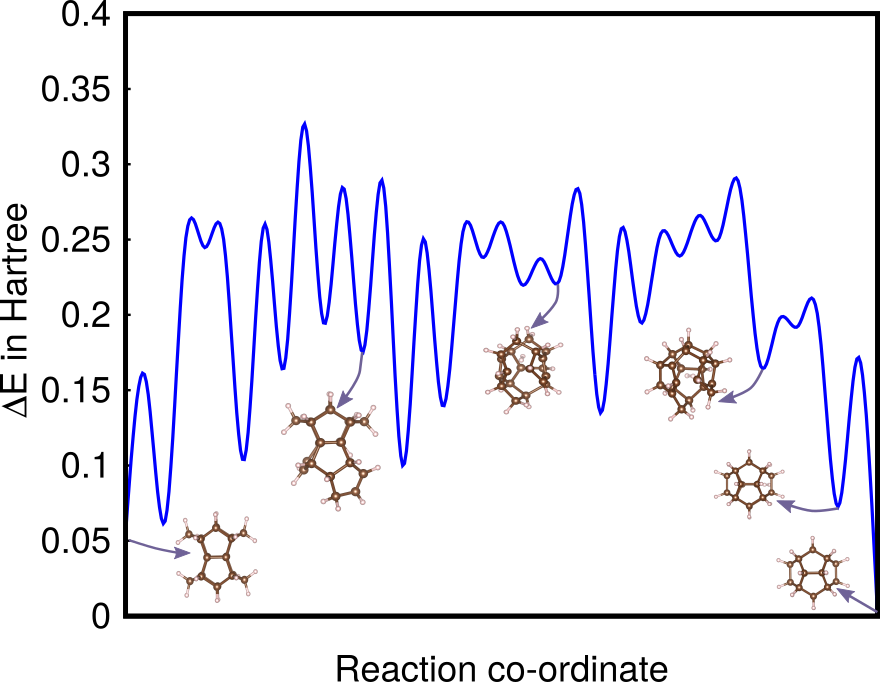}
\caption{ Typical low-barrier pathway of \ce{C20H20} from the pagodane to the dodecahedron configuration. } 
\label{fig:C20H20dodeca}
\end{figure}

Using the same approach  as for \ce{C60} we also obtained the reaction pathway of \ce{C20H20} from its pagodane configuration to its dodecahedron ground state.
The reaction pathway, obtained from a data base of 900 saddle points, is shown in Fig~\ref{fig:C20H20dodeca}. In contrast to \ce{C60} 
there is no systematic lowering of the energy of the intermediate structures as one 
approaches the ground state. This reflects the fact that there is indeed no one step 
synthesis recipe to obtain the dodecahedron from the pagodane structure~\cite{Pagodesynthesis}. 
The experimental synthesis procedure consist of 11 single steps, 
each of which involves other ingredients that modify the PES and are therefore expected to give the necessary driving force towards the ground state
~\cite{fessner1987pagodane,prakash1988stable,fessner1987dodecahedranes}.

In summary, we have introduced a bias that is invariant under atomic index permutations and that can target a single well defined configuration 
as the final configuration of a chemical reaction or physical transformation.
In this way we can overcome the index mapping problem that would require an exponentially large number of applications of 
standard two sided saddle point search methods, since one has to try out in principle all possible index permutations. 
The forces arising from the bias, by construction, do not depend on the indexing of the atoms. We have thus reduced the combinatorial 
atomic indexing problem, that has an exponential scaling, to a global minimization problem on a biased PES involving an indexing invariant 
penalty function. For suitably chosen weights the biased PES has the form of 
a structure seeker and the optimization problem can be solved quite rapidly in practice by global optimization 
methods such as Minima Hopping. 
In contrast to standard order parameters that can drive the system only into certain regions of a low dimensional order 
parameter space, this method can pull the system towards a single configuration in the 
full $3N$ dimensional configuration space. Collective variables are in general system 
specific and finding them can be non-trivial. The penalty function we proposed is universal 
and can be applied to any reaction or transformation in molecules or clusters.
We expect that this method will give atomistic insight into complex reaction pathways  
found for instance in catalysis as well as complex phase and shape transformations in 
nano-particles. Such processes play for instance an important role in non-classical nucleation~\cite{Nonclassical} where the particles undergo transformations similar to the ones studied here for the Lennard Jones cluster. Since the cost of our method is comparable to the cost of a Minima Hopping structure prediction run, 
the method can be applied to systems of the same size. This means that on the density functional level systems with up to 
about 100 atoms can be studied. With cheaper methods such as machine learning based force fields much larger systems will become accessible.
Permutationally invariant distances could also be useful in other reaction path search 
methods where the atomic index mapping is a limiting factor.

\begin{acknowledgments}
This work was done within the NCCR MARVEL.
Computer resources were provided at CSCS under project s707 and at the Scicore 
computing center of the university of Basel. We also thank Dr. Luigi Genovese for support in adapting the BigDFT code. 
\end{acknowledgments}

\bibliographystyle{apsrev4-1}
\bibliography{RPath}

\end{document}